\documentclass[aps,onecolumn,showpacs,nofootinbib,showkeys]{revtex4-2}
\usepackage[margin=2.5 cm]{geometry}

\RequirePackage[T1]{fontenc}
 
\usepackage{epsfig,graphicx,amsmath,amssymb,bm}
\RequirePackage{mathptmx}      
\usepackage{mathrsfs}
\usepackage{amssymb}
\RequirePackage{color}
\usepackage{changes}
\usepackage{slashed}
\usepackage{multirow}
\usepackage{scalerel}
\usepackage{tikz-feynman}
\usepackage{textcomp}
\usepackage{subcaption}
\usepackage{float}
\RequirePackage{hyperref}
\hypersetup{
    linktocpage,
    colorlinks,
    citecolor=blue,
    filecolor=black,
    linkcolor=blue,
    urlcolor=blue,
}
\usepackage{nccmath}

\begin{document}

\title{
Decays of radially excited vector mesons $\rho'$ and $\omega'$ in the extended NJL model
}


\author{M.K. Volkov$^{1}$}\email{volkov@theor.jinr.ru}
\author{A.B. Arbuzov $^{1}$}\email{arbuzov@theor.jinr.ru}
\author{A.A. Pivovarov $^{1}$}\email{tex$\_$k@mail.ru}
\author{K. Nurlan$^{1,2,3}$}\email{nurlan@theor.jinr.ru}

\affiliation{$^1$ Bogoliubov Laboratory of Theoretical Physics, JINR, 
                 141980 Dubna, Russia \\
                $^2$ The Institute of Nuclear Physics, Almaty, 050032, Kazakhstan \\
                $^3$ L. N. Gumilyov Eurasian National University, Astana, 010008, Kazakhstan}   


\begin{abstract}
The decay widths of $\rho' \to \pi\pi, \omega\pi, \rho\eta$, $\omega' \to \rho\pi, \omega\eta$ and $\pi' \to \rho\pi$ are recalculated within the extended NJL quark model. It is shown that the mass of $\rho'(1450)$ meson in the derived in the NJL model agrees with the one given in Particle Data Group tables but significantly deviates from the recent result of CMD-3 collaboration.
Comparisons with earlier theoretical results are presented.
%
%
\end{abstract}


\maketitle


\section{\label{Intro}Introduction}

Recently, the results of the CMD-3 collaboration experiments at the VEPP-2000 collider at low energies (up to 1.2~GeV) were published~ \cite{CMD-3:2023alj}. New values for the cross--section of electron-positron annihilation as a function of energy were obtained in these experiments. These results are very important for calculating the effect of hadronic vacuum polarization and subsequently for estimating the contribution of the latter to the muon anomalous magnetic moment~\cite{Ignatov:2023wma}. In the process of analyzing the $e^+e^-\to\pi^+\pi^-$ process cross section, new fairly accurate values for the mass and width of the $\rho'$ meson ($\rho'\equiv\rho( 1450)$) are obtained: $M_{\rho'} = 1226.22 \pm 24.76$~MeV and $\Gamma_{\rho'} = 272.97 \pm 45.53$~MeV, respectively. In most models, this meson is considered as the first radially excited state of the $\rho$ meson. The new values of the $\rho'$ meson parameters differ significantly from those given in the tables of Particle Data Group (PDG)~\cite{ParticleDataGroup:2022pth}: $M_{\rho(1450)} = 1465 \pm 25$~MeV and $ \Gamma_{\rho(1450)} = 400 \pm 60$~MeV. Note that the CMD-3 collaboration fitted the parameters of the $\rho'$ meson using data at energies less than its mass. However, the smallness of the obtained errors in these parameters indicates the importance of the $\rho'$ meson contribution to the $e^+e^-\to\pi^+\pi^-$ process and in the region of the $\rho$ meson peak. This situation stimulates interest in the theoretical and experimental study of processes involving the $\rho'$ meson and its properties. 
  

The extended Nambu--Jona-Lasinio (NJL) model allows one to describe the properties and interactions of both the ground and first radially excited states of light mesons \cite{Volkov:1996fk, Volkov:1999yi, Volkov:2005kw, Volkov:2017arr}.
In this paper, we present the results of new calculations for the 
$\rho'$ and $\omega'$ main decay widths: $\rho' \to \pi\pi, \omega\pi, \rho\eta$, $\omega' \to \rho\pi, \omega\eta$ and $\pi' \to \rho\pi$. 
From a theoretical point of view, the decays of radially excited mesons $\rho'$ and $\omega'$ were previously considered in~\cite{Volkov:1996fk, Volkov:1999yi, Volkov:2005kw, Volkov:2017arr}. 
The decays of radially excited mesons have been poorly studied experimentally, but such studies are planned in experiments at VEPP-2000, Belle II, BES III, the Super $c - \tau$ factory, etc.
Direct experimental measurements of these decays can be carried out in current and future experiments that will allow a better understanding of the nature of these mesons and, in particular, will indirectly provide a new analysis of the consistency of the latest CMD-3 results.

\section{QUARK-MESON LAGRANGIAN OF THE EXTENDED NJL MODEL}
The $U(3) \times U(3)$ extended NJL~\cite{Volkov:1996fk, Volkov:1999yi, Volkov:2005kw,Volkov:2017arr} model will be used here to describe the listed above decays. In this model, the first radially excited mesons are described by introducing the simplest form factor in polynomial form of the
second degree in the relative momentum of quarks
$F \left(k_{\perp}^2 \right) = c f\left(k_{\perp}^{2}\right)\Theta(\Lambda^2_3 - k_{\perp}^2)$, where 
$f\left(k_{\perp}^{2}\right) = \left(1 + d k_{\perp}^{2}\right)$ \cite{Volkov:1996fk, Volkov:2005kw}. Here the constant $c$ affects the determination of the meson mass. Function $f\left(k_{\perp}^{2}\right)$ is relevant for description of quark-meson interactions. The numerical coefficient $d$ (slope parameter) is uniquely fixed from the condition of invariance of the quark condensate after taking into account radially excited states, and it depends only on the quark composition of the corresponding meson \cite{Volkov:1999yi, Volkov:2005kw}.
As a result, the quark-meson Lagrangian of the strong interactions of vector, axial-vector and pseudoscalar mesons in the ground and first radially excited states in the extended NJL model takes the form \cite{Volkov:2005kw,Volkov:2017arr}
\begin{eqnarray}
	\label{Lagrangian}
		\Delta L_{int} & = &
		\bar{q} \biggl[ 
		i \gamma^{5} \sum_{j = \pm} \lambda_{j}^{\pi} \left(a_{\pi}{\pi}^{j} + b_{\pi}{\pi'}^{j}\right) +
		+\frac{1}{2} \gamma_{\mu} \sum_{j = \pm} \lambda_{j}^{\rho} \left(a_{\rho}\rho^{j}_{\mu} + b_{\rho}\rho'^{j}_{\mu} \right) \nonumber \\ 
		&& 
		+ \frac{1}{2} \gamma^{\mu} \lambda^{\omega} \left(a_{\rho}\omega_{\mu} + b_{\rho}\omega'_{\mu} \right)	
		+ \frac{1}{2} \gamma_{\mu} \gamma_{5} \sum_{j=\pm} \lambda_{j}^{\rho} a_{a_1}{a_1}^{j}_{\mu}
		+ i\gamma_{5} \sum_{i = u, s} \lambda_{i} a^{i}_{\eta}\eta
		\biggl]q,
\end{eqnarray}
where $q$ and $\bar{q}$ are $u$, $d$, and $s$ quark fields with constituent masses $m_{u} \approx m_{d} = 270$~MeV, $m_{s} = 420$~MeV, and $\lambda$ are certain linear combinations of the Gell-Mann matrices,
\begin{eqnarray}
\label{verteces1}
	a_{M} = \frac{1}{\sin(2\theta_{M}^{0})}\left[g_{M}\sin(\theta_{M} + \theta_{M}^{0}) +
	g'_{M}f_{M}(k_{\perp}^{2})\sin(\theta_{M} - \theta_{M}^{0})\right], \nonumber\\
	b_{M} = \frac{-1}{\sin(2\theta_{M}^{0})}\left[g_{M}\cos(\theta_{M} + \theta_{M}^{0}) +
	g'_{M}f_{M}(k_{\perp}^{2})\cos(\theta_{M} - \theta_{M}^{0})\right],
\end{eqnarray}
where $M$ indicates the corresponding meson, the mixing angles are given in Table~\ref{tab_mixing}. The mixing angles for $\pi$ mesons are $\theta_\pi \approx \theta_{\pi}^0$, so $a_\pi \approx g_\pi$ can be used for the ground states of these mesons. The angles for $\pi$ and $\rho$  mesons were determined from the masses of the corresponding ground and first radially excited states \cite{Volkov:1996fk}. The mixing angle $\theta_{a_1}$ can not be determined in this way if we take the experimental value given in PDG ($M_{a_1} = 1230$~MeV) as the $a_1$ meson mass since this value differs from the theoretical one~($M_{a_1}^2 = M_{\rho}^2 + 6 m_u^2$) \cite{Volkov:2005kw}. Therefore, the mixing angle $\theta_{a_1}$ can be considered as an arbitrary parameter. In the work \cite{Volkov:1997dd}, within the extended NIL model, this angle was chosen as $\theta_{a_1} = \theta_\rho$. Here we consider it more reasonable to choose this angle based on the experimental decay width $\rho^0 \to \pi^+ \pi^- $, as explained in Appendix \ref{app_1}.
 
\begin{table}[h!]
\begin{center}
\begin{tabular}{cccc}
\hline
   & $\pi$ & $\rho$ & $a_1$ \\
\hline
$\theta_M$	& $59.48^{\circ}$	&  $81.80^{\circ}$  & $96.0^{\circ}$  \\
$\theta^0_M$	& $59.12^{\circ}$	& $61.50^{\circ}$  & $61.50^{\circ}$  \\
\hline
\end{tabular}
\end{center}
\caption{Mixing angle values of mesons in the ground and first radially excited states.}
\label{tab_mixing}
\end{table}

For the $\eta$ mesons, the factor $a$ takes a slightly different form. This is due to the fact that in this case four states $\eta, \eta', \eta(1295)$ and $\eta(1475)$ are mixed~\cite{Volkov:1999yi, Volkov:2005kw, Volkov:2017arr}:
        \begin{eqnarray}
            a^{u}_{\eta} & = & 0.71 g_{\eta^{u}} + 0.11 g'_{\eta^{u}} f_{uu}(k_{\perp}^{2}), \nonumber\\
            a^{s}_{\eta} & = & 0.62 g_{\eta^{s}} + 0.06 g'_{\eta^{s}} f_{ss}(k_{\perp}^{2}).
        \end{eqnarray}

The quark-meson coupling constants have the form
\begin{eqnarray}
	\label{Couplings}
 g_{\pi} = g_{\eta^{u}}=\left(\frac{4}{Z_{\pi}}I_{20}\right)^{-1/2}, \quad
\, g'_{\pi}=g'_{\eta^{u}} =  \left(4 I_{20}^{f^{2}}\right)^{-1/2}, \nonumber\\
\, g_{\eta^{s}}=\left(\frac{4}{Z_{\eta^s}}I_{02}\right)^{-1/2}, \quad
\, g'_{\eta^{s}} =  \left(4 I_{02}^{f^{2}}\right)^{-1/2}, \nonumber\\
g_{\rho} =\left(\frac{2}{3}I_{20}\right)^{-1/2}, \quad
\, g'_{\rho} =\left(\frac{2}{3}I_{20}^{f^{2}}\right)^{-1/2}.
\end{eqnarray}
where $Z_{\pi}$, $Z_{\eta^{s}}$ are additional renormalization constants appearing 
in transitions between pseudoscalar and axial-vector mesons~\cite{Volkov:2005kw, Volkov:2017arr}.

Integrals appearing in the quark loops are
\begin{eqnarray}
	I_{n_{1}n_{2}}^{f^{m}} =
	-i\frac{N_{c}}{(2\pi)^{4}}\int\frac{f^{m}(k^2_{\perp})}{(m_{u}^{2} - k^2)^{n_{1}}(m_{s}^{2} - k^2)^{n_{2}}}\Theta(\Lambda_{3}^{2} - k^2_{\perp})
	\mathrm{d}^{4}k.
\end{eqnarray}
where $\Lambda_3=1030$~MeV is the three dimensional cutoff parameter~\cite{Volkov:2005kw}.

\section{AMPLITUDES AND decay WIDTHS}

\begin{figure*}[t]
 \centering
   \centering
    \begin{tikzpicture}
     \begin{feynman}
      \vertex (a) {\(\rho' \)};
      \vertex [dot, right=1.6cm of a] (b) {};
      \vertex [dot, above right=2cm of b] (c) {};
      \vertex [dot, below right=2cm of b] (d) {};
      \vertex [right=1.6cm of c] (g) {\(\omega \)};
      \vertex [right=1.6cm of d] (f) {\(\pi \)};
      \diagram* {
        (a) -- [] (b),
        (b) -- [fermion] (c),
        (c) -- [fermion] (d),
        (d) -- [fermion] (b),  
        (c) -- [] (g),         
        (d) -- [] (f),
      };
     \end{feynman}
    \end{tikzpicture}
   \caption{Triangular quark diagram describing the decay $\rho' \to \omega \pi$
   }
 \label{diagram1}
\end{figure*}
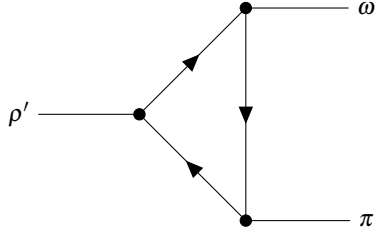%

The decay of $\rho' \to \omega \pi$ is described by a triangular diagram with a triangular quark loop presented in Figure~\ref{diagram1}. When calculating this diagram, the quark loop integral is expanded in terms of the momentum of the external fields, and only the divergent parts are retained \cite{Volkov:2017arr}. As a result, for the $\rho' \to \omega \pi$ decay we obtain the following amplitude
\begin{eqnarray}
\label{amplitude_1}
\mathcal{M}(\rho' \to \omega\pi) = 4 m_u I^{\rho' \omega \pi}_{30} \varepsilon_{\mu\nu\lambda\delta} p_\lambda^{\omega} p_\delta^{\pi} e_\mu(\rho') e^{*}_\nu(\omega),
\end{eqnarray} 
where $e_\mu(\rho')$ and $e^{*}_\nu(\omega)$ are the polarization vectors.

The integral obtained from the triangular quark loop has the form
\begin{eqnarray}
\label{integral}
&& I_{n_1n_2}^{M M'...}(m_{u}, m_{s}) = -i\frac{N_{c}}{(2\pi)^{4}} 
 \int\frac{a(k_{\perp}^{2})...b(k_{\perp}^{2})...}{(m_{u}^{2} - k^2)^{n_1}(m_{s}^{2} - k^2)^{n_2}}
\Theta(\Lambda_{3}^{2} - \vec{k}^2) \mathrm{d}^{4}k,
\end{eqnarray}
where $a(k_{\perp}^{2})$ and $b(k_{\perp}^{2})$ are the coefficients for different mesons defined in (\ref{verteces1}). The decay amplitude of $\omega' \to \rho \pi$ is obtained from (\ref{amplitude_1}) by replacing the vertices $\rho' \to \omega'$ and $\omega \to \rho$. By a similar replacement of the vertices of the corresponding mesons, one can obtain the amplitudes for $\rho' \to \rho\eta$ and $\omega' \to \omega\eta$ decays. 
The decay width $\rho' \to \omega \pi$ can be calculated using the formula
	\begin{eqnarray}
	\label{width_1}
        \Gamma(\rho' \to \omega \pi) =
        \frac{\sqrt{E^2_{\pi} - M^2_{\pi}}}{24 \pi M^2_{\rho'}} \, {\mid \mathcal{M}(\rho' \to \omega\pi) \mid}^2,
	\end{eqnarray}
where $E_{\pi} = (M^2_{\rho'}+M^2_{\pi} - M^2_{\omega}) / 2M_{\rho'}$ is the pion energy in the rest of the $\rho'$ meson. The meson masses are taken from PDG~\cite{ParticleDataGroup:2022pth}.

The decay $\rho' \to \pi\pi$ in the extended NIL model is described by the width
	\begin{eqnarray}
	\label{width_2}
        \Gamma(\rho' \to \pi \pi) =
        \frac{(M^2_{\rho'} - 4 M^2_{\pi})^{3/2}}{48 \pi M^2_{\rho'}} \, {\left[ C_{\rho'} g_\rho Z_\pi \left(1 - \frac{C_{\rho}}{C_{\rho'}} \frac{4m^2_u}{M^2_{a_1}} I^{\rho' a_1}_{20} \right) \right]}^2,
	\end{eqnarray}
where $C_{\rho} = 0.94$ and $C_{\rho'} = 0.33$ \cite{Volkov:2017arr}. 



The strong decay of a radially excited meson $\pi' \to \rho\pi$ in the NJL model is described by the amplitude
\begin{eqnarray}
\label{amplitude_3}
\mathcal{M}(\pi' \to \rho\pi) = 4 I^{\pi' \rho \pi}_{20} \left[ 1 - \frac{2m^2_u}{M^2_{a_1}} \frac{I^{\pi' a_1 \rho}_{20} I^{a_1 \pi}_{20}}{I^{\pi' \rho \pi}_{20}} \right] (p_{\pi'} + p_{\pi})_\mu e^{*}_\mu(\rho),
\end{eqnarray} 
where $p_{\pi'}$ and $p_{\pi}$ are meson momenta. 


The calculated strong decay widths of radially excited mesons $\rho'$, $\omega'$ and $\pi'$ in framework of the extended NJL model and comparison with the results of works \cite{Gutsche:2008qq, Barnes:1996ff, Barnes:2002mu,Mengesha:2013xab} are presented in Table~\ref{tab_width}. The error of the used model is estimated at 15\% \cite{Volkov:2017arr}. The decays $\rho' \to \omega \pi$, $\rho' \to \pi\pi$, $\omega' \to \rho\pi$ and $\pi' \to \rho\pi$ were previously described in the NJL model in paper \cite{Volkov:1997dd}. That work differs from the present one in that, when taking into account the pion production through the $a_1$ meson at the outer ends, an approximation was used leading to a reduction in the constant $Z_{\pi}$ and the choice of the mixing angle for the axial-vector meson $\theta_{a_1} = \theta_\rho$.

\begin{table}[h!]
\begin{center}
\begin{tabular}{ccccc}
\hline
Decay mode &  NJL & \cite{Gutsche:2008qq} &  \cite{Barnes:1996ff,Barnes:2002mu}  &   \cite{Mengesha:2013xab}  \\
\hline
$\rho' \to \omega \pi$ 		& 125		& 165            &122            & 98.6  \\
$\rho' \to \rho \eta$ 		        & 21			& 19            &25           & --  \\
$\rho' \to \pi\pi$			        & 98	                 & 108             &74               & --  \\
$\omega' \to \rho\pi$		        & 309		 &422                  & 328	& 141   \\
$\omega' \to \omega\eta$		& $9$	 		 & 9                 & 12		        & --   \\
$\pi' \to \rho\pi$				& 102 			 & 257              &209         	&  --  \\
\hline
\end{tabular}
\end{center}
\caption{Decay widths in MeV}
\label{tab_width}
\end{table} 

\section{CONCLUSION}

In the present work, the main strong decay widths of the first radially excited vector mesons $\rho'$ and $\omega'$ are calculated. In Table~\ref{tab_width}, the comparison of the obtained results in the framework of the NJL model with the theoretical results of the works \cite{Gutsche:2008qq, Barnes:1996ff, Barnes:2002mu,Mengesha:2013xab} is given. In the work \cite{Gutsche:2008qq}, the calculations were made using the chiral Lagrangians obtained on the basis of the Chiral Perturbation Theory. In the works \cite{Barnes:1996ff, Barnes:2002mu}, the wave functions of the nonrelativistic quark model were applied to estimate the amplitudes and widths of the decays. In the work \cite{Mengesha:2013xab}, the decays $\rho' \to \omega\pi$ and $\omega' \to \rho\pi$ were described by using the Bethe-Salpeter equation (BSE). For the decays of anomalous type our results are in agreement with the data of the theoretical works mentioned above. In the decay $\pi' \to \rho\pi$, the width obtained in the present work turned out to be half as large as the results of other theoretical works \cite{Gutsche:2008qq, Barnes:1996ff, Barnes:2002mu,Mengesha:2013xab}. In our opinion, such inconsistency is a consequence of taking into account $\pi - a_1$ transition in the final $\pi$ meson.

In the case of the $\rho'$ meson decay, we used the standard value of the mass $M_{\rho'} = 1465 \pm 25$ MeV given in PDG \cite{ParticleDataGroup:2022pth}. In the experiment CMD-3 at VEPP-2000 carried out recently, the new value of the mass $M_{\rho'} = 1226.22 \pm 24.76$ MeV \cite{CMD-3:2023alj} was obtained, and it significantly differs from the standard value given in PDG. This value of the $\rho'$ meson mass is questionable. The most important point is that with such value of the $\rho'$ meson mass the decay $\rho' \to \rho\eta$ observed in a series of experiments is forbidden. Besides, in the extended NJL model, theoretical predictions for the masses $M_{\pi'}=1300$ MeV and $M_{\rho'}=1470$ MeV \cite{Volkov:1996fk, Volkov:1999yi} are in the agreement with the data from PDG \cite{ParticleDataGroup:2022pth}. Also, in the extended NJL model, when using the mass $M_{\rho'} = 1226$ MeV, it is impossible to diagonalise the Lagrangian of the ground and first radially excited states and calculate some model parameters \cite{Volkov:2005kw,Volkov:2017arr}. Using such mass when keeping the values of other parameters would obviuously lead to a significant change in the predictions for the decay widths considered in the present paper. That is why, we consider it important to carry out more accurate measurements of the decays of radially excited vector mesons.

\appendix
\section{The definition of the $a_1$ meson mixing angle}
\label{app_1}
The process $\rho^0 \to \pi^+ \pi^-$ in the standard NJL model \cite{Volkov:1986zb} is described with the amplitude 
\begin{eqnarray}
    \mathcal{M}(\rho^0 \to \pi^+ \pi^-) = g_{\rho} e_{\mu}(\rho) \left(p_{+} - p_{-}\right)^{\mu},
\end{eqnarray}
where $p = p_{+} + p_{-}$ is the $\rho$ meson momentum.

By using the experimental value of the width of this process ($\Gamma(\rho^0 \to \pi^+ \pi^-)_{exp} = 149$ MeV) one can fix the value of the constant $g_{\rho} = 6$. This constant is used to determine the main model parameters --- mass of the $u$ quark and the cut-off parameter \cite{Volkov:2017arr}.

The extended NJL model applied in the present paper allows one to take into account apart from the ground  meson states also the first radially excited ones. However, the transition to the extended NJL model leads to a change in the amplitude of the decay $\rho^0 \to \pi^+ \pi^-$:
\begin{eqnarray}
    \mathcal{M}(\rho^0 \to \pi^+ \pi^-) = 4 I_{20}^{\rho \pi \pi} \left[1 - 4\frac{I_{20}^{\rho a_1 \pi} I_{20}^{a_1 \pi}}{I_{20}^{\rho \pi \pi}} \frac{m_u^2}{M_{a_1}^2}\right] e_{\mu}(\rho) \left(p_{+} - p_{-}\right)^{\mu},
\end{eqnarray}
where the integrals $I_{20}^{\rho \pi \pi}, I_{20}^{\rho a_1 \pi}$ and $I_{20}^{a_1 \pi}$ are defined in (\ref{integral}).

As a result, new parameters, mixing angles of the ground and first radially excited meson states, appear in the amplitude. The mixing angles of the $\rho$ and $\pi$ mesons were defined using the experimental values for the masses respective ground and first radially excited states \cite{Volkov:2017arr}. However, the attempt to determine the value of the mixing angle of the $a_1$ meson leads to the appearance of imaginary parts. Thus, this angle was used previously chosen to be equal to the $\rho$ meson mixing angle. As a result, when keeping the values of the parameters determined in the standard NJL model, the theoretical width of the given process turns out to be different from the value obtained in the standard NJL model and, consequently, from the experimental value ($\Gamma(\rho^0 \to \pi^+ \pi^-) = 132$ MeV). However, this angle can be determined from the requirement that the transition to the extended NJL model does not lead to the deviation of the theoretical width of the decay $\rho^0 \to \pi^+ \pi^-$ from the experimental data. Thus, for this angle one can obtain the value $\theta_{a_1} = 96^{\circ}$.

\subsection*{Acknowledgments}
This work was supported by the Science Committee of the Ministry of Science and Higher Education of the Republic of Kazakhstan (Grant no. BR21881941).


\end{document}